\documentclass{aa}
\usepackage{psfig}
\begin{document}

	\thesaurus{08    % A&A Section 6: stars
              (02.13.1;  % MHD,
               06.19.2;  % (Sun:) solar wind,
               08.06.2;  % Stars: formation,
               08.13.2;  % Stars: mass-loss,
               09.10.1)} % ISM:jets and outflows.
	\title{Outflows from magnetic rotators}
	\subtitle{II. Asymptotic structure and collimation}
	\author{T. Lery \inst{1,2},
	J. Heyvaerts \inst{1} , S. Appl \inst{1,3},
	\and 
	C.A. Norman \inst{4}
		}
	\offprints{T. lery}
	\institute {Observatoire de Strasbourg
 	   11 rue de l'Universit\'e
	   67000 Strasbourg, France
	\and 
	   Department of Physics,
	   Queen's University, Kingston,
	   Ontario, K7L 3N6, Canada
	\and
	   Institut fuer Angewandte Mathematik, 
	   Universitaet Heidelberg, 
	   Im Neuenheimer Feld 293, D-69120 Heidelberg, Germany
	\and 
           Space Telescope Science Institute
           and Johns Hopkins University
           3700 San Martin Drive, Baltimore, MD 21218, USA
		}
\date{Received 10 August 1998; accepted 18 May 1999}

\titlerunning{Outflows from magnetic rotators. II}
\authorrunning{T.Lery et al.}

\maketitle

\begin{abstract}

The asymptotic structure of outflows from rotating magnetized objects
confined by a uniform external pressure is calculated. The flow is 
assumed to be perfect MHD, polytropic, axisymmetric and stationary.
The well known associated first integrals together with the confining 
external pressure, which is taken to be independent of the distance 
to the source, determine the asymptotic structure. The integrals 
are provided by solving the flow physics for the base within the 
framework of the model developed in Paper I (Lery et al. 1998), which 
assumes conical geometry below the fast mode surface, and ensures the 
Alfv\'en regularity condition. Far from the source, the outflow 
collimate cylindrically. Slow (i.e. with small rotation parameter 
$\omega$) rigid rotators give rise to diffuse electric current 
distribution in the asymptotic region. They are dominated by gas 
pressure. Fast rigid rotators have a core-envelope structure in 
which a current carrying core is surrounded by an essentially current 
free region where the azimuthal magnetic field dominates. The total 
asymptotic poloidal current carried away decreases steadily with the 
external pressure. A sizeable finite current remains present for fast 
rotators even at exceedingly small, but still finite, pressure. 

\end{abstract}

\section{Introduction}

Jets from young stellar objects (YSO) and active galactic nuclei (AGN) are 
most likely launched magnetically. Various approaches have been used to 
describe the stationary configuration of magnetically collimating winds 
governed by the Grad-Shafranov equation (e.g. Lery et al. \cite{lery1} 
hereafter Paper I, and references therein). Magnetized rotating MHD winds 
can be accelerated from an accretion disk (``Disk wind'', Blandford \& Payne 
\cite{bp}, Pelletier \& Pudritz \cite{pellpud}), at the disk-magnetosphere 
boundary (``X-winds'', Shu et al. \cite{shul}, \cite{shu94}, \cite{shu97}) 
or directly from the star itself by combined pressure and magneto-centrifugal 
forces (``Stellar wind'', Weber \& Davis \cite{wd}, MacGregor \cite{macg} and 
reference therein). In order to make the system of equation more 
tractable angular self-similarity has been often employed for non rotating 
magnetospheres (Tsinganos \& Sauty \cite{tsin1}), and outflows from spherical 
rotating objects (Sauty \& Tsinganos \cite{sautytsing}, Trussoni et al. 
(\cite{trussoni}), Tsinganos et Trussoni \cite{tsin2}). Cylindrical 
self-similarity has also been used for magnetized jets (Chan \& Henriksen 
\cite{chanh}), and spherical self-similarity for disk winds (Blandford \& 
Payne \cite{bp}, Henriksen \& Valls-Gabaud \cite{henrik}, Fiege \& Henriksen 
\cite{fiege}, Contopoulos \& Lovelace \cite{conto}, Ferreira \& Pelletier 
\cite{ferr1}, Ferreira \cite{ferr3}, Ostriker \cite{ostriker}, Lery et al.
\cite{lhf}). However this assumption presents boundary condition restrictions.
Several numerical simulations, such as Ouyed \& Pudritz \cite{ouyedp}, have 
been made in order to understand the formation of magnetized jets from 
keplerian discs, but due to computational limitations only a few cases have 
been studied. One should also note that pure hydrodynamic collimation could 
be effective at producing jets (Frank \& Mellema \cite{FM}). High velocity 
outflows from YSO and AGN are observed to be highly collimated. Heyvaerts 
\& Norman (\cite{HN}) have discussed how streamlines asymptotically develop 
in  winds with different properties without considering confinement by any 
external medium (see also Heyvaerts \cite{HN2}). They have shown that winds 
which carry a non-vanishing Poynting flux and poloidal current to infinity 
must contain a cylindrically collimated core, whereas other winds focus 
parabolically. Li et al. (\cite{liCB}) showed how the formation of weakly 
collimated, conical flows depends on the shape of the poloidal field near 
the Alfv\'en surface. The transition from weakly collimated flows to highly
collimated jets has been also studied by Sauty \& Tsinganos 
(\cite{sautytsing}).

\paragraph{Close to the Source}
In Paper I, a model for the stationary structure of the inner part 
of the flow has been proposed based on the assumption that the 
magnetic surfaces possess a shape which is a priori 
known inside the fast critical surface. As a first approximation magnetic 
surfaces were taken to be cones. Unlike the Weber-Davis type models, the 
balance of forces perpendicular to the magnetic surfaces is taken into 
account on the Alfv\'enic surface through the Alfv\'en regularity condition. 
This, together with  the criticality conditions determine  the three unknown
constants of the motion, that are conserved along magnetic surfaces $a$, 
namely the specific energy $E(a)$, the specific angular momentum $L(a)$ 
and the mass to magnetic flux ratio $\alpha(a)$. Once these first integrals 
are determined, the asymptotic cylindrically collimated flow is uniquely 
determined. The solutions are parameterized by  the angular velocity of the 
magnetic field lines $\Omega(a)$, the specific entropy at the base $Q(a)$ 
and by the mass flux to magnetic flux ratio on the polar axis $\alpha_0$. 
According to the rotation parameter $\omega=\frac{\Omega r_A}{v_{PA}}$ the 
objects can be classified as slow ($\omega << 1$), fast 
($(\frac{3}{2})^{3/2}-\omega << 1$) or intermediate (other values of 
$\omega$ ranging between 0 and $(\frac{3}{2})^{3/2}$) rotators. Critical 
surfaces are nearly spherical for slow rotators, but become strongly 
distorted for rapid rotators, giving rise to important gradients of density
and velocity that should consequently effect asymptotic quantities. This 
simplified model makes it possible to investigate the structure of outflows
far from the magnetized rotator source without the need for self-similar 
assumptions. The price to pay for that is that the model does not give an 
exact, but only an approximate, solution because the transfield equation 
is not solved everywhere, but only at a few special places.

How outflows behave in the asymptotic region constitutes the subject
of the study of this paper which also focuses on the study of the 
asymptotic electric current and addresses the question of the asymptotic 
collimation of  the different classes of rotators found in Paper I.
In this paper, we present a model where the collimated jet is assumed to 
be in pressure equilibrium with an external medium whose properties are 
independent of the distance to the source. The question of the asymptotic 
electric current is a major concern for jets that will be investigated.

We structure our paper as follows: In \S2, we present the equations 
governing the asymptotic equilibrium, and their boundary conditions. 
Solutions and parameter studies are presented in \S3. We derive 
analytical and numerical solutions for the asymptotic electric 
current in \S4. Finally, in\S5, we discuss the implications of our 
analysis and we summarize our results in \S 6.

\section{The Analytical Model}

\subsection{The Jet Base}

%%%%%%%%%%%%%%%%%%%%%%%%%%%%
\begin{figure}
\psfig{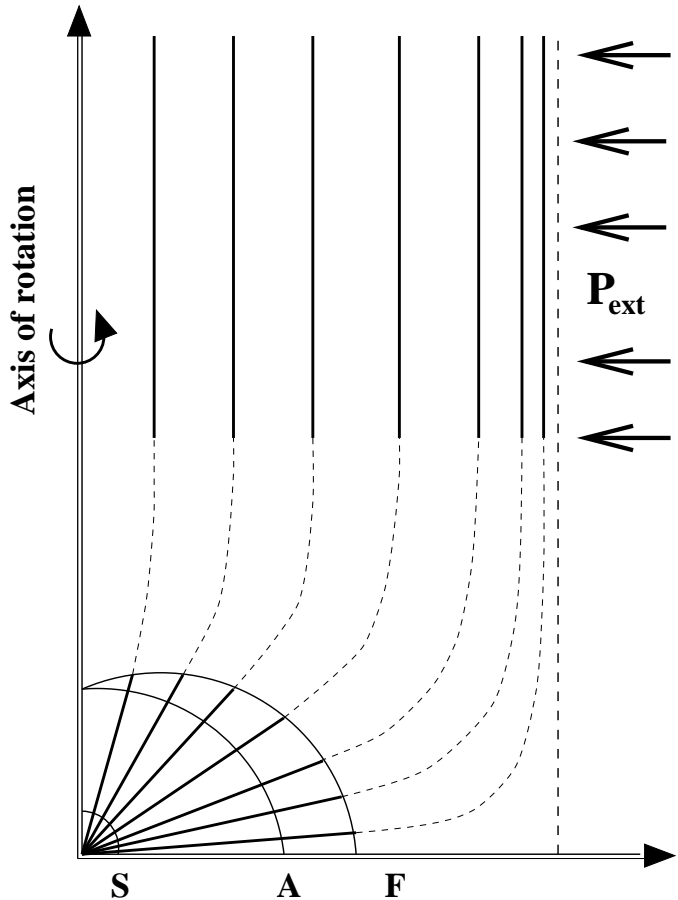}
\caption[ ]{Schematic representation of 
the magnetic structure of the model.
 The magnetic surfaces projected in the 
poloidal plane are conical within the fast magnetosonic surface
and connect to the cylindrical asymptotic region.
In this part, the outflow is surrounded
 by an external confining medium.
S, A and F denote the slow magnetosonic, 
the Alfv\'en and the fast magnetosonic surfaces, respectively.}
\label{fig1}
\end{figure}
%%%%%%%%%%%%%%%%%%%%%%%%%%%%
Magnetic surfaces in the inner region are assumed to be conical up to the 
fast magnetosonic surface. The flow eventually becomes cylindrical due to 
confinement by uniform external pressure (Fig.~\ref{fig1}). Henceforth we 
work in the cylindrical coordinate system ($r$,$\phi$,$z$) whose axis 
coincides with the symmetry-axis. Each flux surface is labeled by the 
flux function $a(r,z)$ proportional to the magnetic flux through a circle 
centered on the axis passing at point $r$, $z$. The physical flux is 
$2 \pi a$. The equatorial value of this function is A. In addition to 
being steady and axially symmetric, we further constrain the mass density 
$\rho$ of the flow to be related to the gas pressure $P$ by a polytropic 
equation of state $P =  Q(a) \rho^{\gamma}$.
We refer to $Q(a)$ as the "specific entropy" of the flow (though it would be 
related to it only for adiabatic flows). The constant $\gamma$ is the 
polytropic index which is considered to be constant. The specific entropy 
$Q(a)$, the angular velocity of the magnetic field lines $\Omega(a)$ and 
the mass to magnetic flux ratio $\alpha(a)$ are constant along any flux 
surface $a$, and entirely determine the outflow in the conical region. 
The total energy $E(a)$, the total angular momentum $L(a)$, and the ratio 
of matter to magnetic flux $\alpha(a)$, expressed through the density at 
the Alfv\'en point $\rho_A(a)$, are also conserved along the flux surfaces 
and follow from the regularity of the solution at the Alfv\'enic and fast 
and slow magnetosonic surface. 

\subsection{The Intermediate Zone}
The MHD flow in regions causally disconnected from the base region has no 
back-reaction on its properties in this region. It is known that this  
causally disconnected  region starts at the so-called fast limiting 
characteristic (Tsinganos et al. \cite{tsin3}), which is usually situated 
downflow from the fast magnetosonic surface. The flow in the causally 
disconnected region has no influence on the values assumed by the three 
first integrals which are not a-priori known. The flow between the fast 
mode critical surface and the fast mode limiting separatrix would have 
some influence on their determination if the shape of magnetic surfaces 
between the source object and the fast limiting separatrix were 
self-consistently calculated by solving exactly the transfield equation.
However, in the present simplified and fixed base geometry, these first 
integrals are entirely determined from sub-fast surface regions and there is 
no need, for their determination, nor for that of the asymptotic structure, 
to calculate the geometry of magnetic surfaces in the intermediate region 
between the fast critical surface and the fast mode limiting separatrix. The 
latter is anyway presumably located not much further away from it and we do 
not expect large geometrical changes as compared to the base region. The lack 
of complete self-consistency of our model is therefore mainly contained in 
our assumption of base conical geometry. The fact that the shape of surfaces
is not calculated downflow from the fast surface in regions were field 
curvature is still present, i.e. where poloidal field lines are dashed 
in Fig.~\ref{fig1}, does not add any supplementary inaccuracy. 

The shape of the magnetic surfaces shown in Fig.~\ref{fig1} are similar, 
to some extent, to shapes obtained by some previous studies already 
performed for a self-consistent calculation of the shape of the poloidal 
field lines. Trussoni et al. (\cite{trussoni}) have prescribed similar 
types of the magnetic surfaces and then integrated the MHD equations from 
the base all the way to infinity. Such case corresponds to meridionally 
self-similar MHD outflows with a non-constant polytropic index $\gamma$.
Sauty and Tsinganos (\cite{sautytsing}) have also calculated the shape 
of magnetic field lines by deducing them and integrating the MHD equations 
from the base to large distances.

\subsection{The Asymptotic Structure}
The asymptotic structure of the flow is determined by
the Bernoulli equation and the force balance in the 
direction perpendicular to the field (transfield equation) 
in terms of the constants of motion $Q, \Omega, E, L, \alpha$. 

\paragraph{The Bernoulli Equation}

Let us define $L$ as the total angular momentum, $\rho_A$ as the mass 
density at the Alfv\'en critical point and $G(r,z)$ as the gravitational 
potential. The Bernoulli equation can be given by
\begin{eqnarray}
{1\over 2}{\frac{\alpha^2 \nabla a^2} 
{\rho^2 r^2}}
=
E(a) 
- G(r,z)
- {\gamma \over \gamma - 1} 
Q \rho^{\gamma - 1}
\nonumber \\
+\rho\Omega
\frac{L-r^2\Omega}{\rho_A-\rho}
-\frac{1}{2}
\left(
\frac{L}{r}+\frac{\rho}{r}
\frac{L-r^2\Omega}{\rho_A-\rho}
\right)^2 .
\label{equa1}
\end{eqnarray}
This equation can be simplified in the asymptotic region,
i.e. $z$ going to $\infty$, so that gravity becomes negligible.
In the cylindrical case, $\nabla a$ is replaced by $da/dr$. Since we are 
far from the Alfv\'en surface in the asymptotic region, the density of 
the flow $\rho$ must be smaller than the Alfv\'enic density $\rho_A$.
Moreover we can consider $r$ to be larger that $r_A$.
Using the assumptions  $r\gg r_A$ and $\rho_A\gg\rho$, the 
last two terms of the Eq.~(\ref{equa1}) become
\begin{equation}
-\frac{\rho\Omega^2r^2}{\rho_A}
-\frac{1}{2}
\frac{r_A^4\Omega^2}{r^2}
\left[1-\frac{\rho r^2}{\rho_A r_A^2}
\right]^2 .
\end{equation}
This is equivalent to
\begin{equation}
-\frac{\rho\Omega^2r^2}{\rho_A}
-\frac{\rho\Omega^2r^2}{\rho_A}
\left[
\frac{1}{2}\left(\frac{\rho_A r_A^2}{\rho r^2}\right)\frac{r_A^2}{r^2}
-\frac{r_A^2}{r^2}
+\frac{1}{2}\frac{\rho}{\rho_A}
\right] .
\end{equation}
The last two terms in brackets are negligible w.r.t. unity.
Moreover when $z$ tends to infinity, $\rho r^2$ is bounded.
Therefore the parenthesis of the first term inside the brackets 
is also bounded, and, if $r$ were to approach infinity, the first 
term in bracket would also be negligible with respect to unity. 
We assume that, even though $r$ approaches a finite limit,  
($r_A/r$) becomes asymptotically small enough for this  first 
term in bracket to also become negligible. Thus the last two 
terms of the Bernoulli equation can be approximated at infinity by
$-\rho\Omega^2r^2/\rho_A$.
Hence, the Bernoulli equation becomes
\begin{equation}
{1\over 2}\left({{\alpha \over {\rho r}} 
{d a\over d r}}\right)^2 =
E - {\gamma \over \left(\gamma - 1 \right)} 
Q \rho^{\left(\gamma - 1 \right)}
-{{\Omega^2 r^2 \rho} \over {\mu_0 \alpha^2}}.
\label{bernoulli2}
\end{equation}

\paragraph{The Transfield Equation}
The force balance perpendicular to the field can be written 
in cylindrical coordinates as
\begin{eqnarray}
&{{\alpha}\over{\rho r}}\left(
{{\partial}\over{\partial z}} {{\alpha}\over{\rho r}}
{{\partial a}\over{\partial z}}
+ {{\partial}\over{\partial r}} {{\alpha}\over{\rho r}}
{{\partial a}\over{\partial r}}
\right)
- {{1}\over{\mu_0 \rho r}}
\left(
{{\partial}\over{\partial z}} {{1}\over{r}}{{\partial a}
\over{\partial z}}
+ {{\partial}\over{\partial r}} {{1}\over{r}}{{\partial a}
\over{\partial r}}
\right)
\nonumber \\
&=
E' - {{Q' \rho^{\gamma -1} }\over{\gamma - 1}}
+ {{\alpha'}\over{\alpha}} \ \ {{\mu_0 \alpha^2 \rho}\over{r^2}} \ \ 
{{(L - r^2 \Omega)^2}\over{(\mu_0 \alpha^2 - \rho)^2}} 
\nonumber \\
& 
- {{\rho}\over{r^2}} \ \ 
 {{(L'-r^2 \Omega')(L - r^2 \Omega)}\over{\mu_0 \alpha^2 - \rho}}
- {{L  L'}\over{r^2}}.
\label{Grad}
\end{eqnarray}
Primes denote derivatives with respect to $a$, i.e. $E'= dE/da$. 
Similarly using the same asymptotic assumptions in the transfield 
equation, the centrifugal force $\rho v_\phi^2/r$ can be neglected with 
respect to ``hoop stress'' $B_{\phi}^2/\mu_0 r$ since
\begin{equation}
\frac{\rho v_{\phi}^2}{r} =
\frac{B_{\phi}^2}{\mu_0 r} 
\left( \frac{r_A^2}{r^2} \right)
\left( \frac{\rho_A r_A^2}{\rho r^2} \right)
\left( 1 - \frac{\rho r^2}{\rho_A r_A^2} \right)^2 .
\end{equation}
Hence due to the simplifications, several force densities vanish in the 
one-dimensional form of the transfield equation that is then given by
\begin{equation}
\frac{1}{2}
 {d \over da} 
\left(\frac{\alpha}{\rho r}{da \over dr}\right)^2
=
E^\prime
-\frac{Q^\prime \rho^{\gamma-1}}{\gamma-1}
+\frac{\rho r^2 \Omega^2}{\mu_0 \alpha^2}
\left(\frac{\alpha^\prime}{\alpha}-\frac{\Omega^\prime}{\Omega}\right).
\label{trans1}
\end{equation}
Subtracting the transfield equation (\ref{trans1}) from the derivative 
with respect to $a$ of the Bernoulli equation (\ref{bernoulli2}), 
one can simplify the transfield equation that becomes
\begin{equation}
r^2 {d \over da} \left(Q \rho^{\gamma}\right)
+ {1\over {2}}
{d\over da } 
\left({\Omega^2 r^4 \rho^2} \over {\mu_0 \alpha^2}\right)= 0.
\label{trans2}
\end{equation}
Thus Eqs.~(\ref{bernoulli2}) and (\ref{trans2}) describe
the asymptotic equilibrium structure of a magnetized jet with the present
assumptions. To study the equilibrium of this jet taking into 
account an external ambient pressure, one needs to specify the 
relevant boundary condition at the jet's edge.

\paragraph{The Cylindrical Collimation}
It is possible to show that the asymptotic problem with 
non-vanishing external pressure does not accept solutions 
where $r$ goes to infinity on any magnetic field line. 
Indeed, if so, ${da\over dr}$ would vanish at the edge of the jet.
As a consequence $B_p=\frac{1}{r}{da\over dr}$ would go to zero 
at the outer edge and the toroidal part of the magnetic field 
would asymptotically reduce to
\begin{equation}
B_\phi =\mu_0 \alpha \frac{\rho}{\rho_A-\rho}\frac{L-r^2\Omega}{r}
\approx -\frac{\Omega }{\alpha}\frac{\rho r^2}{r} .
\end{equation}
If $\rho r^2$ were to diverge at the edge, the Bernoulli equation 
(\ref{equa1})
would be violated, since the left hand side term  is always positive,
and the ninth term that is negative and the largest one 
in absolute value could not be balanced by other terms.
It would also be so if the Alfv\'en radius were to become infinite. 
This proves that $B_{\phi}$ and $B_{p}$ should vanish if $r$ were 
to approach infinity. If so, the boundary condition reduces to 
$P_{ext} = P_{gas} = Q \rho_b^{\gamma}$.
The density at the outer edge $\rho_b$  would then be finite, 
and $\rho r^2$ would diverge which violates the Bernoulli 
equation as shown above. This proves that {\em the confining pressure
limits the jet to a finite radius as $z \rightarrow \infty$
and therefore ensure an asymptotically cylindrical structure}.

\paragraph{An Upper Limit for the Axial Density}
Let us see now that the physics of the flow in the inner region 
close to the source constrains the maximum value of the asymptotic 
mass density on the polar axis, $\rho_0$, and therefore also the
total mass flux for a given magnetic flux. Indeed, on the axis the 
Bernoulli equation reduces to
\begin{equation}
\left(v_{\rm P}^2/2\right) = E - \gamma
Q \rho_0^{\gamma-1}/(\gamma-1).
\end{equation}
which yields an upper limit to the axial density (the limit corresponding 
to a vanishing asymptotic poloidal velocity),
\begin{equation}
\rho_{0} \le \left(\frac{\gamma-1}{\gamma}
\frac{E}{Q}\right)
^{\frac{1}{\gamma-1}}.
\label{rhozero}
\end{equation}
In the inner region of the flow close to the source the energy has been 
calculated (see paper I). For slow rotators (that corresponds to a
rotation parameter $\omega << 1$) energy is given by
$E=A^2/2 \mu_0^2\alpha^2 R_A^4$,
with the Alfv\'en spherical radius equal to
$R_A=\left(C_1\mu_0 A^2/2\right)^{1/(2\sqrt{2}+4)}$,
where $C_1$ is the constant of integration of the transfield equation
that has been defined analytically in Paper I (Eq.~(75)) only as 
a function of the input parameters. Combining the two last equations 
with Eq.~(\ref{rhozero}) the maximum density becomes in this case
\begin{equation}
\rho_{0,{\rm max}} = \left(\frac{\gamma-1}{\gamma}\frac{A^2}
{2 \mu_0^2  \alpha^2 Q}
\left(\frac{2}{C_1\mu_0 A^2}\right)^{\frac{2}{\sqrt{2}+2}}
\right)^{\frac{1}{\gamma-1}}.
\end{equation}
In the vanishing rotation case, it is then possible to find an analytical 
definition of the maximum density on the axis in the asymptotic region 
allowed by the input parameters defining the emitting source properties.
This also shows that the slow rotator limiting density essentially depends 
on the specific entropy $Q$. For fast rotators and using Eq.~(99) of 
Paper I that gives energy, the limiting mass density becomes
\begin{equation}
\rho_{0,{\rm max}} = \left(\frac{\gamma-1}{\gamma}
\frac{3}{2 Q}\frac{A \Omega^2}{\mu_0 \alpha}
\right)^{\frac{1}{\gamma-1}}.
\end{equation}
This limit now depends on the entropy, but also on angular velocity and mass 
to magnetic flux ratio. The latter parameter is related to mass loss rate on 
the axis and therefore determine the axial value of density at the Alfv\'en 
point. Then an increase of $\alpha_0$ naturally reduces this limit. Thus, 
once given the properties of the source in our model ($Q(a)$, $\Omega(a)$, 
$\alpha_0$), the asymptotic axial density possesses an upper limit.

\paragraph{The Boundary Condition}
We further assume the flow to be in pressure equilibrium with an external 
medium whose pressure is constant. Equilibrium at the jet boundary is 
expressed by 
\begin{equation}
P_{\rm ext} = Q\rho^\gamma + (B_{\rm P}^2 + B_\phi^2)/(2 \mu_0),
\end{equation}
with the magnetic contributions  
\begin{eqnarray}
\frac{B_\phi^2}{2\mu_0} &=& \frac{\rho^2 r^2 \Omega^2}{2 \rho_A}, \\
\label{PBphi}
\frac{B_{\rm P}^2}{2\mu_0} &=& \frac{\rho^2}{\rho_A}
\left( E - \frac{\gamma}{\gamma-1}Q\rho^{\gamma-1}
-\frac{\Omega^2 r^2 \rho}{\rho_A} \right ).
\label{PBp}
\end{eqnarray}
It has been used that at the outer boundary $\rho << \rho_A$ and $r >> r_A$.
The pressure of the external medium may have a thermal and a magnetic
contribution, too.In the case of a finite external pressure the jet radius 
remains finite.

Thus the asymptotic forms of the transfield and the Bernoulli equations and 
the pressure balance at the jet outer edge constitute the set of equations 
describing the asymptotic structure of the pressure-confined jet.
The three first integrals of the motion  $E(a)$, $L(a)$ and $\alpha(a)$ 
are obtained from the inner part of the flow (Paper I). At infinity the 
only free parameter is the external pressure $P_{ext}$.

\section{Numerical Analysis}

\subsection{Numerical Procedure \label{sect24}}

For the numerical calculations, the Bernoulli and the transfield 
equations have been reformulated as two ODEs for the radial position, $r$,
and the density, $\rho$, as a function of the flux surfaces $a$. 
The Bernoulli equation can be written as 
\begin{equation}
\frac{d r}{d a}  =\frac{ \alpha}
{\rho r \sqrt{2}\sqrt{E-\frac{\gamma}{\gamma-1}
Q \rho^{\gamma-1}-\frac{\Omega^2 r^2 \rho}{\mu_0 \alpha^2}}},
\label{B1}
\end{equation}
and the transfield equation can be written as
\begin{eqnarray}
\left(\mu_0  \gamma Q\rho^{\gamma-2} +
\frac{r^2 \Omega^2}{\mu_0 \alpha^2}\right)
\frac{1}{\rho}\frac{d \rho}{d a}
& + & \left(\frac {2r\Omega^2 }{\mu_0 \alpha^2}\right)
\frac{d r}{d a} 
\nonumber \\
=\frac{r^2\Omega^2}{\mu_0 \alpha^3}
\frac{d \alpha}{d a}-
\mu_0\rho^{\gamma-2}
\frac{d Q}{d a}
 &-& \frac{r^2\Omega^2}{\mu_0 \alpha^2}
\frac{d \Omega}{d a} .
\label{T1}
\end{eqnarray}
These two equations can now be written symbolically as
\begin{eqnarray}
\frac{d r}{d a} &=& f_r(r, \rho, E, Q, \alpha,...), \\
\frac{d \rho}{d a} &=& f_{\rho}(r, \rho, E, Q, \alpha,...),
\end{eqnarray}
where $f_r$ is the r.h.s. of Eq.~(\ref{B1}) and $f_{\rho}$ has a more 
complex form that can be easily derived from Eqs.~(\ref{T1}) and (\ref{B1}).
We use a standard initial condition integrator for stiff systems
of first order ODEs. We prescribe the axial density rather than 
the external pressure, which is ultimately deduced from the solution.
As in paper I, dimensionless quantities $\bar\Omega$, $\bar\alpha_0$ 
and $\bar Q$ will respectively be used for $\Omega$, $\alpha_0$ and 
$Q$ in order to simplify numerical investigations. The reference units 
(in CGS) are $\rho_{ref} = 70 p.cm^{-3}$, $r_{ref} = 10^{15}cm$,
$v_{ref} = 10^7cm.s^{-1}$.

We will first consider the case of magnetized winds originating from
objects with constant rotation and entropy. The effect of the four 
parameters is studied in the next subsections, followed by a specific 
application to the TTauri star BP Tau.

\subsection{Variations of the Rotation} 
First we study the effect of the dimensionless angular velocity, 
$\bar{\Omega}(a) = \Omega_\ast$ of the central object. The other 
parameters remain constant.
%%%%%%%%%%%%%%%%%%%%%%%%%%%%%%%%%%%%
\begin{figure}[htbp]
\psfig{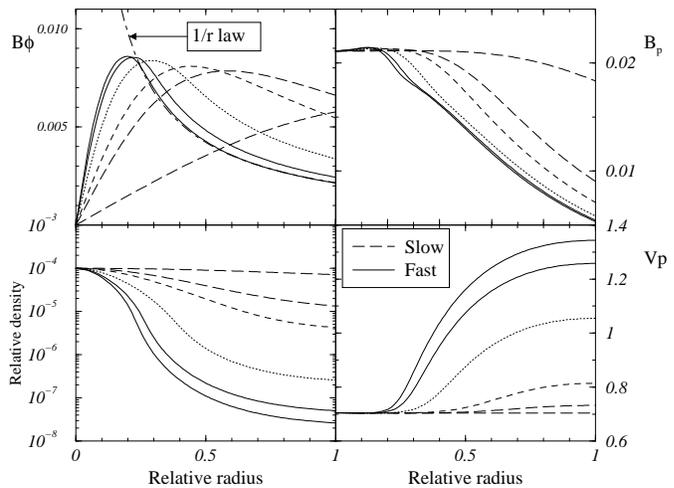}
\caption[ ]{
Rotational Effects:
The  azimuthal $B_\phi$ and poloidal $B_P$ magnetic field components
(upper left and right panels respectively), density $\rho$ and poloidal 
velocity $v_{\rm P}$ (lower left and right panels) are plotted as 
functions of relative radius $r_{\rm rel}$ for different values of 
$\bar \Omega$ ($\bar\Omega=2$(dashed lines),$4,6,8,10,11$(solid lines)).
The corresponding values of $\omega$ are 0.4,1.1,1.3,1.65,1.75 and 1.8.
($\bar Q=2.1$ and $\bar\alpha_0=1.7$)
}
\label{fig4}
\end{figure}
%%%%%%%%%%%%%%%%%%%%%%%%%%%%%%%%%%%
As in paper I, we find that the variations of the angular velocity
have major effects on solutions as can be seen on Fig.~\ref{fig4}
that shows the components of the magnetic field, the density and the 
poloidal velocity for different rotation rates. $\bar\Omega$ varies 
from $2$ to $11$ ($Q_\ast = 2.1$, $\bar\alpha_0 = 1.7$).
For a slow rotator (heavy dashed lines in Fig.~\ref{fig4}), the toroidal
field increases nearly linearly, while it possesses a maximum within the 
jet for fast rotators. We find that $B_\phi\propto 1/r$ while 
$B_p \propto 1/r^2$. The two solutions respectively corresponds to a 
diffuse current with nearly constant current density, and to a centrally 
peaked current, surrounded by a current-free envelope. A similar behavior 
has been discussed by Appl \& Camenzind (\cite{applcam}) for relativistic 
jets, according to which the jet configurations had been referred to as 
diffuse and sharp pinch. Important variations with respect to rotation 
can also be seen in the other physical quantities. In particular the 
poloidal velocity is highest in the envelope, though variations remain 
within a factor of two. In the very fast rotator limit the density falls 
off dramatically in the envelope where the gas pressure becomes negligible 
compared to the magnetic pressure. This is clearly shown in Fig.~\ref{fig5}
where the total pressure is represented with magnetic and gas pressures.
%%%%%%%%%%%%%%%%%%%%%%%%%%%%%%%%%%%%
\begin{figure}[htbp]
\psfig{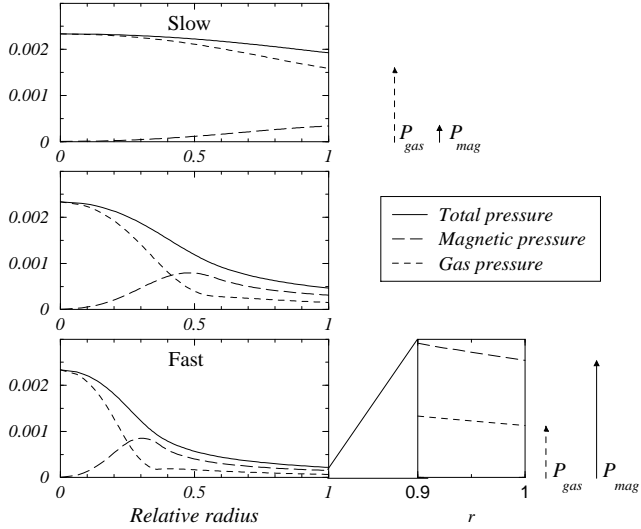}
\caption[ ]{Rotational Effects:
Comparison of pressure profiles for slow, fast and very fast rotators. 
Total, magnetic ($P_{mag}$) and gas ($P_{gas}$) pressure respectively 
correspond to solid, dashed and long dashed lines. The rotation 
parameter $\omega$ is equal to 0.4, 1.3 and 1.8 top to bottom respectively. 
}
\label{fig5}
\end{figure}
%%%%%%%%%%%%%%%%%%%%%%%%%%%%%%%%%%%
{\em For slow rotators, the gas pressure dominates everywhere in the outflow,  
while for fast rotators the magnetic pressure dominates in the envelope.} 
In the latter case, most of the outer pressure at the outer edge is 
supported by magnetic field and not by gas pressure.

%%%%%%%%%%%%%%%%%%%%%%%%%%%%%%%%%%%%
\begin{figure}[htbp]
\psfig{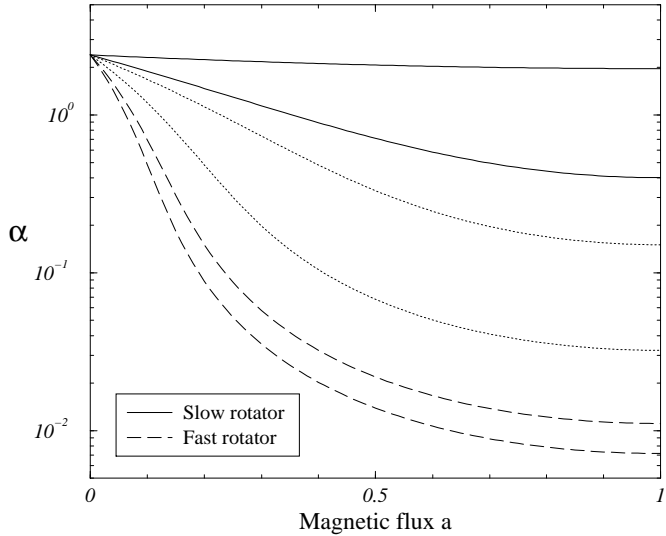}
\caption[ ]{
Rotational Effects:
Plot of the mass to magnetic flux ratio $\bar\alpha$ with 
respect to the relative magnetic flux $a_*=\frac{a}{A}$
for various constant rotation rates $\bar\Omega = 2..11$. 
Solid lines correspond to
slow rotators, and long dashed lines to outflows with 
large $\bar\Omega$. All the other parameters and boundary 
conditions are kept the same.
}
\label{fig6}
\end{figure}
%%%%%%%%%%%%%%%%%%%%%%%%%%%%%%%%%%%
In Fig.~\ref{fig6}, the mass to magnetic flux 
ratio $\bar\alpha$ is represented as a function of the 
relative magnetic flux for various angular velocities.
Since the central part of the outflow is denser
for fast rotators, the mass flux is very large
in this region and very concentrated around the axis.
It means that {\em most of the matter is flowing
along the polar axis for fast rotators}.

\subsection{Mass Loss Rate Effects}

%%%%%%%%%%%%%%%%%%%%%%%%%%%%%%%%%%%%
\begin{figure}[htbp]
\psfig{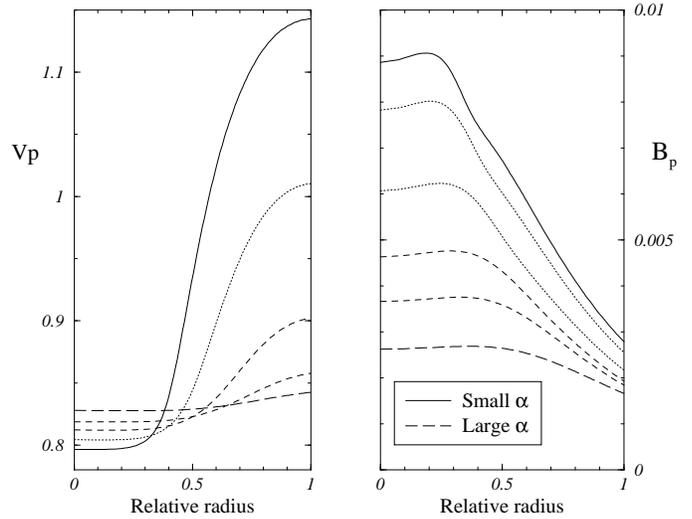}
\caption[ ]{
Mass Loss Rate Effects:
Plots of poloidal components of velocity $v_p$ (left panel) and 
magnetic field $B_p$ (right panel) for various values of 
the mass to magnetic flux ratio on the axis $\bar\alpha_0$.
Heavy solid line correspond to small mass loss rates
and heavy dashed line to large ones.
($\bar\alpha_0=0.7,0.8,1,1.4,1.8,2.6$, $\bar\Omega=5$ and 
$\bar Q=2.5$)
}
\label{fig7}
\end{figure}
%%%%%%%%%%%%%%%%%%%%%%%%%%%%%%%%%%%

Another potentially observable quantity is the mass loss rate
particularly interesting since it can be evaluated from observations. 
The parameter related to the mass loss rate is $\bar\alpha_0$. We have 
computed solutions for different $\bar\alpha_0$'s, keeping other 
parameters unchanged. The results are plotted in Fig.~\ref{fig7} where
the poloidal components of the velocity $v_p$ and the magnetic field $B_p$ 
are represented as functions of the relative radius. The input values are 
$\bar\alpha_0=0.7,0.8,1,1.4,1.8,2.6$, $\bar\Omega=5$ and $\bar Q=2.5$.
From this figure we can infer that {\em the maximum momentum is situated in 
the axial part, especially for fast rotators}. Therefore the central part 
of the jet will propagate more easily in the ambient medium than the 
external part. 

When the mass loss rate grows, the outflow is slowed down on the edge and 
accelerated on the axis, while the poloidal magnetic field is also reduced. 
Therefore {\em the jets from the youngest stellar objects, which show the 
largest mass loss rates, should have a faster central core and a slower 
envelope than older ones}. 

The profile of poloidal velocity depends sensitively on $\bar\alpha_0$ 
as well as the core radius that reduces as $\bar\alpha_0$ increases. 
As found in Paper I for the inner conical region, it appears that  
an increase in mass loss rate has a similar effect to a decrease of 
the rotation rate.

\subsection{Thermal Effects}

\begin{figure}[htbp]
\psfig{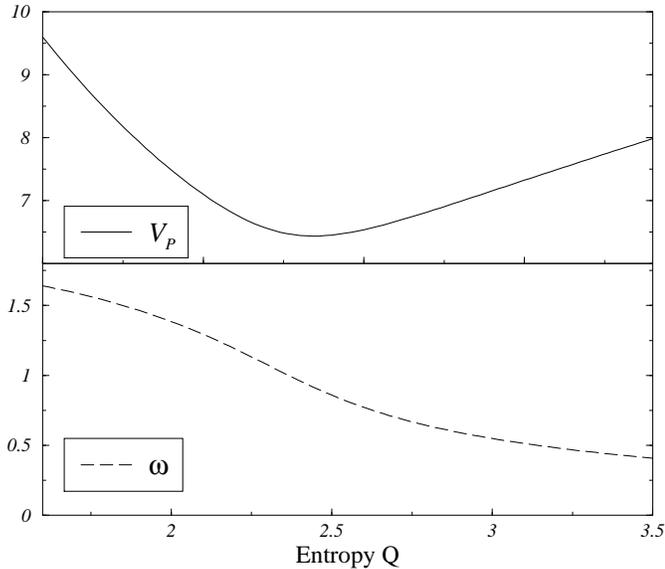}
\caption[ ]{
Thermal Effects:
Poloidal velocity $v_p$ (upper panel) 
and rotation parameter $\omega$ (lower panel)
are plotted for $\bar Q$ varying from 1.2 to 3.5
given at the boundary of the outflow.
$\omega$ varies from $0$ to its maximum value. 
$\bar\Omega=\Omega_\ast=5$, $\bar\alpha_0=2.1$ and $\rho_0=10^{-5}$.
}
\label{fig8}
\end{figure}

In this subsection we study the dependence of the outflow properties
on specific entropy $Q_\ast$ on the stellar surface and polytropic 
index $\gamma$. The other parameters are given by 
$\bar\Omega=\Omega_\ast=5$, $\bar\alpha_0=2.1$ and $\rho_0=10^{-5}$.
Fig.~\ref{fig8} shows the poloidal velocity $v_p$ and the rotation 
parameter $\omega$ calculated at the outer edge of the flow as 
functions of the specific entropy $\bar Q$. The whole range in 
$\omega$ can be covered by only varying $\bar Q$. Smaller specific 
entropies correspond to larger $\omega$. For given $\bar\Omega$ and 
$\bar\alpha_0$ the poloidal velocity presents a minimum when the 
rotation parameter $\omega$ is almost equal to unity which corresponds 
to the intermediate class of rotators. Slow rotators are then 
accelerated by an increase of the heating, but not fast rotators.
It has also been found that a change in specific entropy does not 
affect the azimuthal to poloidal magnetic field ratio but just 
produces an increase of magnitudes.

%%%%%%%%%%%%%%%%%%%%%%%%%%%%%%%%%%%%
\begin{figure}[htbp]
\psfig{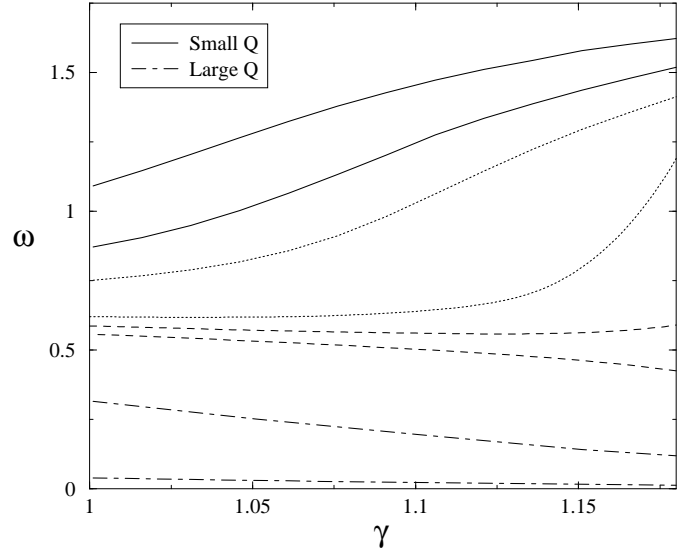}
\caption[ ]{
Thermal Effects:
The rotation parameter $\omega$ given at the outer boundary
as a function of $\gamma$ for different values of $\bar Q$
(from 1 to 3.5)
The smallest values of $\bar Q$ (represented with solid lines) 
increase and reach the largest value of $\omega$, while
larger value of $\bar Q$ (dashed lines) decrease and correspond 
to smaller values of $\omega$.
}
\label{fig9}
\end{figure}

%%%%%%%%%%%%%%%%%%%%%%%%%%%%%%%%%%%
It has been found that the velocity can change by several orders of 
magnitude, between the adiabatic  and isothermal flow, which are the 
narrower under the same conditions.
In Fig.~\ref{fig9}, the combined effects of $\bar Q$ and $\gamma$ 
are represented. The rotation parameter $\omega$ is plotted with 
respect to $\gamma$ for various values of $\bar Q$. Two different 
classes of solutions can be distinguished, solutions for which 
$\omega$ start at small values for $\gamma =1$ and then decrease
and solutions which have large $\omega$ at $\gamma = 1$ and 
then increase. Thus {\em the value of the specific entropy  discriminates 
between fast and slow rotators while larger  $\gamma$'s just tighten 
this distinction}. 
%%%%%%%%%%%%%%%%%%%%%%%%%%%%%%%%%%%%
%%%%%%%%%%%%%%%%%%%%%%%%%%%%%%%%%%%

\subsection{An Example: BP Tau}

%%%%%%%%%%%%%%%%%%%%%%%%%%%%%%%%%%%%
\begin{figure}[htbp]
\psfig{figure=8055.f8,width=\linewidth}
\caption[ ]{
Model for the jet of a TTauri star, BP Tau with 
$\bar\Omega = 1.8$, $\bar\alpha_0 = 0.1$, $\bar Q = 0.05$ 
and a relative central density of $10^{-4}$ ($\omega=1.41$).
Upper panels correspond to the magnetic field (left)
and the velocity components (right). Density and gas pressure 
(right) together with the relative magnetic flux (left) are 
represented in the lowest panels as a function of the relative radius
$r_{rel}=r/r_{jet}$.
}
\label{fig2}
\end{figure}
%%%%%%%%%%%%%%%%%%%%%%%%%%%%%%%%%%%
We now apply the jet model to the TTauri star BP Tau by using its
 properties, given by Bertout et al.(1988), which are
$\dot M_{\ast}=2\times10^{-7} M_{\odot} yr^{-1}$,
$M_{\ast}=0.8 M_{\odot} $, $R_{\ast}=3 R_{\odot}$, $T_{\ast}=9\times10^3$,
$n_p=10^4 cm^{-3}$, and $B{\ast}=1000 G$.
On the base of these values we deduce (see paper I) the corresponding 
input parameters, which are, expressed in terms of the dimensionless 
reference values mentioned in section \ref{sect24},
$\bar{Q}=0.05$, $\bar{\Omega} = 1.8$, $\bar{\alpha_0}= 0.1$
and a relative central density of $10^{-4}$. These parameters allow 
to compute the dimensionless rotation parameter $\omega$ which is found 
to be 1.41 on the equator, and corresponds to the case of a fast rotator. 
Magnetic field and velocity components, density and gas pressure in the 
flow are plotted for BP Tau in Fig.~\ref{fig2} together with the magnetic 
flux represented as a function of the relative radius. The reference units
(in CGS) are $\rho_{ref} = 70 p.cm^{-3}$, $r_{ref} = 10^{15}cm$,
$v_{ref} = 10^7cm.s^{-1}$.

The azimuthal component of the magnetic field dominates the poloidal part 
in the envelope. The poloidal velocity does not show strong variations 
across the jet, and the azimuthal velocity $v_\phi$ is one order of magnitude  
smaller than the poloidal component. {\em The axial region is the 
densest and slowest part of the asymptotic flow}. This dense core region 
is a fraction of the full jet with a minimum of $0.5$ for the fastest case.
This corresponds to a central core of the order of $5\times10^{14}cm$.
Even if the fastest part is the outer one the maximum momentum is 
located around the polar axis. Hence even if the relative velocity
is smaller close to the axis the central region of the flow will 
propagate faster in the ambient medium. It also is found that, only in 
the central part of the asymptotic outflow, does the kinetic energy 
flux dominate over Poynting flux. Thus {\em a large part 
of the magnetic energy has not been transferred to the kinetic 
energy in the case of constant rotation}.

\subsection{Non-constant $\Omega$ and $Q$}

The profiles of the differential angular velocities, that we use
(See Paper I), are defined as follows: $\Omega_0$ corresponds to a 
constant rotation rate across the flow, $\Omega_1$ is a profile varying 
from $\Omega_0$ to zero with a step-like transition, the same for 
$\Omega_3$ but more smoothly, $\Omega_2$ and $\Omega_5$ vary from 
$\Omega_0$ to $\Omega_0/2$ and $1.5\times\Omega_0$ respectively 
and finally $\Omega_4$ follows the differential rotation of a 
Solar-type star. The profiles of specific entropies are as follows: 
$Q_0$ is constant across the flow, $Q_1$ varies from $Q_0$ to 
$Q_0/2$ and $Q_2$ varies from $Q_0$ to zero like $Q_1$ but more 
smoothly in the latter case.

%%%%%%%%%%%%%%%%%%%%%%%%%%%%%%%%%%%%
\begin{figure}[htbp]
\psfig{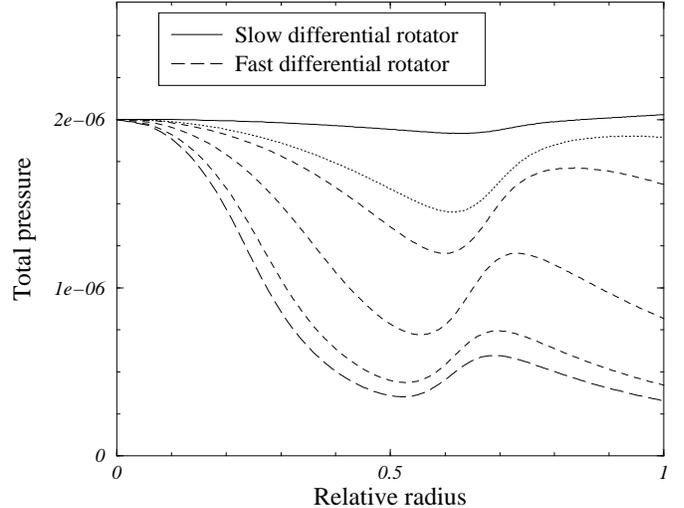}
\caption[ ]{
Differential Rotators:
we plot the total pressure as a function of the relative radius $r_{rel}$. 
Slow rotators are represented with solid lines and fast differential 
rotators with dashed lines. Rotation rate is given by $\Omega_2$.
}
\label{fig12}
\end{figure}
%%%%%%%%%%%%%%%%%%%%%%%%%%%%%%%%%%%

A number of well-collimated outflows  are observed to 
have larger poloidal velocities near the polar axis, and lower 
velocities at the edge of the flow. This motivated us to use a 
profile of type $\Omega_2$ that could reproduce such behaviors.
Fig.~\ref{fig12} represents the total pressure for such 
differential rotators with respect to the relative radius
for a set of central values of the angular velocity. Similarly 
to rigid rotators, the pressure globally decreases from the axis 
to the outer edge, though the gradient of rotation causes a second 
peak of pressure to appear. The latter could correspond to a denser,
slower and wider outflow surrounding the central fast jet.
For slow and intermediate rotators,  the inner pressure
can be of order of the pressure at boundary of the outflow
and the outer part can be as dense and slow as the axial region.
A peak of poloidal velocity accompanies the pressure minimum
which produces a double structure in such outflows, with a dense slow
core surrounded by a faster component at half of the total radius
itself embedded in a slower outer part. Differential rotators can 
then produce jets with a narrow central part with large momentum 
surrounded by a larger outflow with smaller momentum, 
as are jet surrounded by a molecular flow.

It is also found that some outflows like rigid rotators
have smaller axial velocities than at the outer edge or 
present on the contrary a very fast component which 
corresponds to large gradients of the angular velocity.
Differential entropies can also cause such inverted 
asymptotic solutions though with less amplitude in the 
variation of the poloidal velocity.

%%%%%%%%%%%%%%%%%%%%%%%%%%%%%%%%%%%%
\begin{figure}[htbp]
\psfig{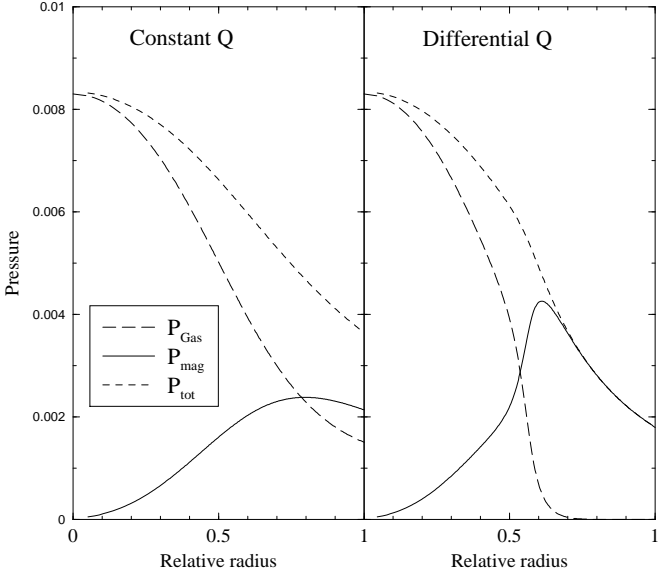}
\caption[ ]{
Comparison of the magnetic (solid), the gas (long dashed) and 
the total (dotted) pressures for constant (left) and differential (right)
specific entropies. Parameters correspond to a fast rotator. 
The angular velocity $\bar\Omega$ is constant.
}
\label{fig15}
\end{figure}
%%%%%%%%%%%%%%%%%%%%%%%%%%%%%%%%
Variations of the entropy profile bring an interesting feature.
We have shown previously that magnetic pressure dominates at the 
outer edge for fast rotators while gas pressure dominates for slow 
rotators. In fact, variations of the specific entropy enhances this 
difference as shown in Fig.~\ref{fig15} where the pressures are plotted
with respect to the relative radius. The total pressure does not
change much as compared to the case of constant entropy 
while the gas to magnetic pressure ratio $\beta$ decreases.
Therefore $\beta$ can become quite low in some regions of the jet.
This can have important consequences on the propagation of the jet
and on the role of ambipolar diffusion (Frank et al. \cite{franketal})
 in the dynamics of large scale YSO jets and outflows.

\section{The Asymptotic Electric Current}

Heyvaerts  and Norman (1989)  have shown that the asymptotic shape
of steady axisymmetric magnetized unconfined outflows is either 
paraboloidal or cylindrical according to whether the electric 
current carried at infinity vanishes or not. As shown in previous 
sections, the uniform confining external pressure causes the flow 
to asymptote to a  cylindrical shape. The aim of this section is 
to study the variation of the poloidal electric current w.r.t.
the confining pressure.

The physical poloidal current through a circle centered on the axis
of symmetry and flowing between this axis and a magnetic surface $a$
is given by $I_{phys}(a) = 2 \pi r B_{\phi}/\mu_0$. The related, 
usually positive quantity $I = -{I_{phys}}/{2\pi}$, which we will 
still refer to as the "poloidal electric current" therefore contains 
entirely equivalent information. The value of the flux surface index, 
$a$, at the equator corresponds to the total magnetic flux enclosed 
in the outflow and is defined by  
$A = \int_{0}^{\pi/2} B R^2 cos{\theta} d\theta$.
We make the magnetic flux dimensionless by dividing it by $A$ and use 
$a_* = a/A$.Let us define the following dimensionless variables
normalized to their value at the Alfv\'en surface,
$x \equiv r^2/r_A^2$ and $y \equiv \rho/\rho_A$,
the thermal parameter $\beta \equiv {2 \gamma Q 
\rho_A^{\gamma-1}}/(\gamma-1) v_{PA}^2$, the 
energy parameter $\epsilon \equiv {{2 E }/{v_{PA}}^2}$,
the gravity parameter $g \equiv {{2 G M}/{r_A v_{PA}}^2}$.
In terms of these variables the poloidal current
can be written, by Eq.~(9) which is valid for
$r \gg r_A$,  as
\begin{equation}
I = \frac{A}{ \mu_0 r_A} \omega x y
\label{I1}
\end{equation}

\subsection{Current and Pressure Balance}

We would like to relate the axial density to the value of the
external pressure. It has been found convenient to first consider 
constant $\Omega(a)$ and  $Q(a)$ to investigate the dependence of 
the electric current on model parameters and on the external 
confining pressure. Using the dimensionless variables $x$ and 
$y$ defined above the equations for the asymptotic structure 
of the flow are written as
\begin{equation}
\left({{d x}  \over {d a_*}}
+{\frac{2 x}{r_A}} {{d r_A} \over {d a_*}}
\right)^2 =
\frac{4}{y^2 
\left(\epsilon - \beta y^{\gamma-1} - 2 \omega^2 x y
\right)}
\label{B2}
\end{equation}
\begin{equation}
\left(\frac{\gamma-1}{\gamma}\right) x r_A^2\, 
\frac{d}{d a_*}
\left(\frac{\beta y^{\gamma}}{r_A^4}\right)
+  \frac{d}{d a_*}
\left(\frac{\omega^2 x^2 y^2}{r_A^2}\right) = 0.
\label{T2}
\end{equation}
Considering that the variation of the position of the Alfv\'en 
point with respect to the relative magnetic flux is small, 
the Bernoulli equation (\ref{B2}) can be written as 
\begin{equation}
\dot x = {d x \over d a_* } =
\frac{2}{y \sqrt{\epsilon - 
\beta y^{\gamma-1} - 2 \omega^2 x y}}.
\end{equation}
Assuming  $\Omega(a)$ and $Q(a)$ to be constant,
the transfield equation similarly reduces to
\begin{equation}
\dot y = {d y \over d a_* } =
- y \left({2 \omega^2 \dot x}
\over
{\left(\gamma-1\right)\beta y^{\gamma-2} + 2 \omega^2 x}
\right) .
\label{TRANSgen}
\end{equation}
This equation integrates as
\begin{equation}
\beta y^{\gamma-1}+2 \omega^2 x  y = C
\label{C}
\end{equation}
where C is an integration constant
 which can be related to the axial density
$y(x=0) = y_0$ as
\begin{equation}
C = \beta y_0^{\gamma-1}
\label{defC}
\end{equation}
Using Eq.~(\ref{C}), Eq.~(\ref{I1}) becomes
\begin{equation}
I = 
\frac{A }{2 r_A \mu_0}
\frac{\beta }{\omega}
\left(y_0^{\gamma-1}-y_b^{\gamma-1}\right)
\label{Ibound}
\end{equation}
where $y_b$ is the density at the outer edge. The condition for 
a vanishing asymptotic current is simply $y_0 = y_b$. In order 
to relate the latter density to the external pressure, the 
transfield equation (\ref{TRANSgen}) can be reformulated in 
differential form and, making use of the definition of its 
integration constant $C$ in Eq.~(\ref{C}), can be cast in 
the equivalent form
\begin{equation}
\dot y =
- \frac{2 \omega^2 y^2\dot x}
{\left(\gamma-2\right)\beta y^{\gamma-1} + C}
\label{TRANS3}
\end{equation}
and the Bernoulli equation is simply
\begin{equation}
\dot x = 2/y \sqrt{\epsilon - C}
\label{BERN2}
\end{equation}
Substituting  $\dot x$  from Eq.~(\ref{BERN2}) in Eq.~(\ref{TRANS3}) 
we obtain 
\begin{equation}
\dot y \left(\left(\gamma-2\right)\beta y^{\gamma-2} + C/y \right) =
- 4 \omega^2/A\sqrt{\epsilon-C}
\label{TRANS4}
\end{equation}
that can be integrated between the axis and the outer edge of the
outflow ($y$ varies from $y_0$ to $y_b$ and $\bar a$ varies from 0 to 1)
\begin{equation}
C \ln{\left(\frac{y_b}{y_0}\right)}
+\frac{\beta (\gamma-2)}{\gamma-1}
\left(y_b^{\gamma-1}-y_0^{\gamma-1}\right)
=-\frac{4\omega^2}{\sqrt{\epsilon-C}} .
\label{inte}
\end{equation}
Note that for this integration the energy has been considered as 
almost constant w.r.t. $a$, {\em i.e.} it does not vary significantly 
across the jet. This approximation is justified numerically.
Pressure balance at the outer edge of the jet gives a 
relation between the external pressure and the axial density
\begin{equation}
\tilde P_{ext}= 
\frac{\gamma-1}{\gamma} \beta y_b^{\gamma} +
\omega^2 x_b y_b^2 + y_b^2 
(\epsilon -  \beta y_b^{\gamma-1}-2 \omega^2 x_b y_b)
\label{pextgen}
\end{equation}
where $\tilde P_{ext}$ is defined by
\begin{equation}
\tilde P_{ext} \equiv \frac{2 \mu_0 r_A^4}{A^2} P_{ext}.
\end{equation}
Then the pressure balance equilibrium reduces to
\begin{equation}
\tilde P_{ext}= 
\frac{\gamma-1}{\gamma} \beta y_b^{\gamma}+
\omega^2 x_b y_b^2  + y_b^2 
(\epsilon - C)
\end{equation}
The system of Eqs.~(\ref{inte}) and (\ref{pextgen})
establishes  a relation between the external 
pressure $\tilde P_{ext}$ and the axial density $y_0$.
Further progress in making it explicit is possible in
 the case of slow and fast rotators, since
simple solutions for $\epsilon$ have been found in paper I. 

\subsubsection{Slow Rotators}
 
Gas pressure is larger than magnetic pressure at the outer boundary
for slow rotators. In this case $\omega \ll 1$, and therefore we have
\begin{equation}
\tilde P_{ext} = \frac{\gamma-1}{\gamma}
\beta y_b^\gamma \gg \epsilon - C
\label{pextyb}
\end{equation}
The energy parameter $\epsilon$ has been calculated for slow 
rotators in Paper I and is given by 
$\epsilon = 1+\beta+3 \omega^2 - {\it g}\label{epsislow}$.
Considering $g$ negligible with respect to $\beta$ 
as a first approximation,
Eq.~(\ref{pextyb}) is given by 
\begin{equation}
y_b =\left(
\frac{\gamma \tilde P_{ext} }{(\gamma-1)\beta}\right)
^{1/\gamma}
\label{ybg}
\end{equation}
Using the definition of $\epsilon$, Eqs.~(\ref{defC})
and (\ref{ybg}), the Eq.~(\ref{inte}) becomes
\begin{eqnarray}
\frac{1}{\gamma}
\ln\left(\frac{\gamma\tilde P_{ext}}{(\gamma-1)\beta y_0^\gamma}\right)
+\left(\frac{\gamma-2}{\gamma-1}\right)
\left( \frac{\gamma\tilde P_{ext}}{(\gamma-1)\beta y_0^\gamma}\right)
^{\frac{\gamma-1}{\gamma}}
& &
\nonumber \\
=  -\frac{4 \omega^2}{\beta y_0^{\gamma-1}
\sqrt{1+\omega^2-\beta(1+ y_0^{\gamma-1})}}
+\frac{\gamma-2}{\gamma-1}
& &
\end{eqnarray}
The second terms of both the left and right hand sides are
 negligible with respect to the first terms and  this equation can
be simplified to
\begin{equation}
\tilde P_{ext}=
\frac{\gamma-1}{\gamma}\beta y_0^\gamma
\exp{-\frac{4 \omega^2\gamma}{\beta y_0^{\gamma-1}
\sqrt{1+\omega^2-\beta(1+ y_0^{\gamma-1})}}}
\label{slow3}
\end{equation}
Since $\beta$ and $\omega$ have been obtained 
by a solution for the inner
part of the flow close to the source,
this equation is a relation between the external 
pressure and the asymptotic axial density.
It is now possible to calculate the poloidal current
as a function of the density on the axis
\begin{equation}
I \approx
\frac{\gamma-1}{2\mu_0 r_A}
\frac{\beta y_0^\gamma}{\omega}
\left(1-
\exp{\left(-\frac{4 \omega^2\gamma}{\beta y_0^{\gamma-1}}
\right)}
\right)
\label{Islow3}
\end{equation}
All other parameters in this formula are given by the
solution in the inner part of the flow close to the source
and by the boundary condition on the axis. So 
Eqs.~(\ref{slow3}) and (\ref{Islow3}) allow to give a 
relation between the external pressure and the total 
poloidal current in the slow rotator case.

\subsubsection{Fast Rotators}

In this case, the gas pressure is negligible with respect 
to the magnetic pressure at the outer boundary. This means that
\begin{equation}
\tilde P_{ext} = y_b^2 \left(\epsilon -C\right)\ll \epsilon - C
\label{pextfr}
\end{equation}
Using Eq.~(\ref{pextgen}) and condition (\ref{pextfr}), 
Eq.~(\ref{inte}) becomes now
\begin{eqnarray}
\frac{C}{2} 
\ln\left(\frac{\tilde P_{ext}}{\epsilon-C}\right)
-\frac{C}{\gamma-1}\ln\left({\frac{C}{\beta}}\right)
& &
\nonumber \\
-\frac{\beta(\gamma-2)}{(\gamma-1)}
\left( \frac{\tilde P_{ext}}{\epsilon-C}\right)
^{\frac{\gamma-1}{2}}
-\frac{\gamma-2}{\gamma-1} C
& =  & -\frac{4 \omega^2}{\sqrt{\epsilon-C}} ,
\end{eqnarray}
which has two simple solutions, when the external pressure becomes small, 
which are $C \sim 0$ and $C \sim \epsilon \approx 3 \omega^{4\over 3}$.
In the latter case the poloidal current is given by
\begin{equation}
I = \frac{A }{2 r_A \mu_0 \omega}
\left(3\omega^{4/3}-\beta y_b^{\gamma-1}\right).
\end{equation}
If either $\beta$ or the relative density at the boundary 
is small compared to unity, it reduces to
\begin{equation}
I = \frac{3}{2}\frac{A }{r_A \mu_0}\omega^{1/3}
\end{equation}
It has been shown in paper I that the Alfv\'en radius grow with 
$\omega$ faster than $\omega^{1/3}$ for fast rotators. Then the 
current decreases as the rotator gets faster. Fast rotators do not  
carry the largest electric current. When $\omega$ approaches its limit
$\left(\frac{3}{2}\right)^{\frac{3}{2}}$ corresponding to the very 
fast rotator the current approaches
\begin{equation}
I = \left(\frac{3}{2}\right)^{3/2}\frac{A}{r_A \mu_0}
\end{equation}
Since $r_A$ has been found in paper I to be proportional to 
$\alpha^{-1/3}$, the total current increases with $\alpha$. 

\subsection{Variation of the Current across the Flow}

Differentiating the current as given by Eq.~(\ref{I1})
with respect to $a$ we find that
\begin{equation}
\frac{d I}{d a_*} =  \frac{2 A \omega}{r_A \mu_0\sqrt{\epsilon-C}}
\left(1+\frac{4}{\gamma}\frac{P_{B_\phi}}{P_{gas}}\right)^{-1}
\end{equation}
Along  the same lines as above, we can study
the slow rotator case where the gas pressure dominates
and the fast rotator case which has a strong magnetic pressure
at the outer edge. In the former case, one finds that
\begin{equation}
\frac{d I}{d a_*} 
 \sim  \frac{2 A \omega}{r_A \mu_0
\sqrt{1+3\omega^2-\beta-\beta y_0^{\gamma-1}}}
\end{equation}
The slope of the variation of the poloidal current tends 
to vanish as $\omega$ decreases. It is proportional to 
$\frac{\omega}{r_A}$. For slow rotators the derivative is 
always positive and then the current is  diffuse in the outflow.
For fast rotators, $C=\beta y_0^{\gamma-1} \ll \epsilon$ 
and one has
\begin{equation}
\frac{d I}{d a_*} 
\sim \frac{ A \omega^{1/3} \gamma}{2\sqrt{3} r_A \mu_0}
\frac{P_{gas}}{P_{B_\phi}}
\end{equation}
The magnitude of the slope $\frac{d I}{d a_*}$ tends to vanish for 
very fast rotators since the gas to magnetic pressure ratio decreases 
with the rotation parameter. In the latter case the slope approaches 
small values more rapidly than in the slow rotator case. 
Another interesting point is that the gas to magnetic 
pressure ratio starts to decrease very rapidly at a 
small value of the distance $r$ to the axis, 
so its derivative goes rapidly to zero across the jet
as a function of radius. This means that all the current must be
enclosed in a core around the axis for fast rotators,
which should carry concentrated electric current.
Thus it has been found analytically that 
{\em slow rotators have a diffuse current while
fast rotators carry concentrated electric current around the axis}.

\subsection{Evolution of the Current along the Flow}

%%%%%%%%%%%%%%%%%%%%%%%%%%%%%%%%%%%%
\begin{figure}[htbp]
\psfig{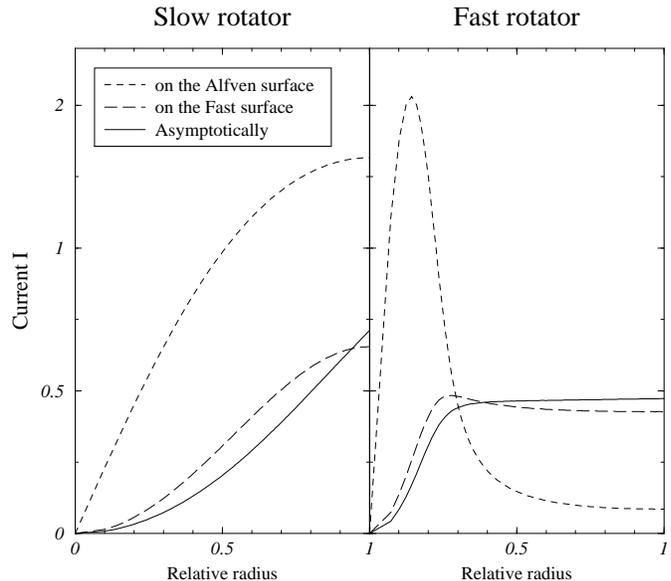}
\caption[ ]{
Variations of the current along the flow for slow 
(left panel) and fast (right panel) rotators as functions 
of the relative radius. The electric current is
calculated in the asymptotic region (solid lines),
at the Alfv\`en surface (dashed lines) and at the fast 
magnetosonic surface (long dashed lines).
}
\label{fig16}
\end{figure}
%%%%%%%%%%%%%%%%%%%%%%%%%%%%%%%%%%%

The variation  of the current along one field line
gives us fruitful informations on the structure of the outflow and
on how the various quantities evolve from the source to infinity.
In Fig.~\ref{fig16}, the currents calculated at the Alfv\'en  
surface, at the fast surface and asymptotically are  plotted
with respect to the relative radius, which is almost linearly 
related to $a_*={a}/A$ with a slope equal to unity. For slow 
rotators, the poloidal electric current globally decreases from 
the Alfv\'en surface to the asymptotic zone, but is constantly 
increasing from the polar axis to the edge of the jet. On the 
other hand the right panel shows that the current at the Alfv\'en 
surface has a peak close to the polar axis for fast rotators.
The current then decreases at larger distances from the axis 
which reveals the existence of a return electric current
flowing around the central electric flow. This illustrates that 
{\em the solutions obtained with this model carry return currents at 
other locations than at the polar axis, contrary  to self-similar 
models}.

\subsection{Influence of Source Properties}

%%%%%%%%%%%%%%%%%%%%%%%%%%%%%%%%%%%%
\begin{figure}[htbp]
\psfig{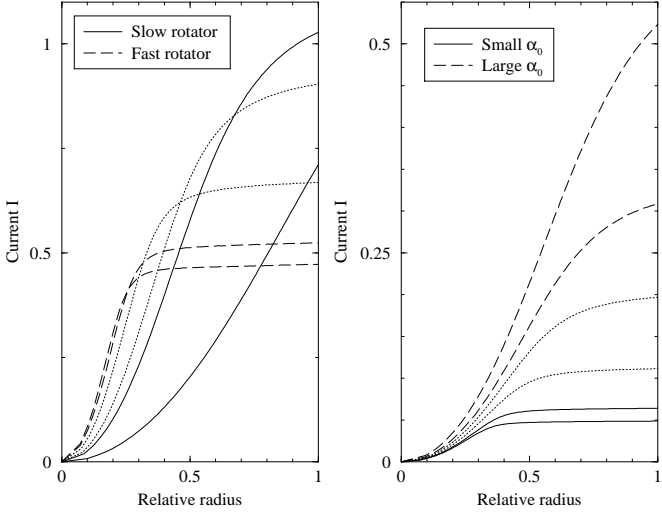}
\caption[ ]{ 
Plots of the asymptotic electric current profile
for different values of constant angular velocity 
$\bar\Omega$ (left panel) and of the mass to magnetic 
flux ratio on the axis $\bar\alpha_0$  (right panel).
Heavy solid lines correspond to slow rotators (left panel) and to
small $\bar\alpha_0$ (right panel), while fast rotators (left panel)
and large $\bar\alpha_0$ are plotted with heavy long dashed lines.
Thin dashed lines correspond to intermediate values. 
}
\label{fig17}
\end{figure}
%%%%%%%%%%%%%%%%%%%%%%%%%%%%%%%%%%

The  parameters whose changes produce the strongest variations
of the solutions are the angular velocity $\bar\Omega$ and 
the mass to magnetic flux ratio on the axis $\bar\alpha_0$. The 
left panel of Fig.~\ref{fig17} represents the variations of the 
asymptotic poloidal current with the relative radius for various 
values of $\bar\Omega$. As the rotation increases, the current 
first increases but soon starts to decrease. Moreover the maximum 
current does not correspond to the largest rotation rate. 
On the right panel of Fig.~\ref{fig17}, the current is plotted for 
different values of $\bar\alpha_0$ for a given $\bar\Omega$.
The larger the mass to magnetic flux ratio, the more the properties
of the outflow resemble those of a slow rotator.
The current profile changes  from a concentrated one, 
signature of a fast rotator, to a more diffuse one
for the largest $\bar\alpha_0$.
The total current increases at the meantime. 
So {\em jets with large mass loss rate also
possess a large current}.

\subsection{Influence of the External Pressure}

%%%%%%%%%%%%%%%%%%%%%%%%%%%%%%%%%%%%
\begin{figure}[htbp]
\psfig{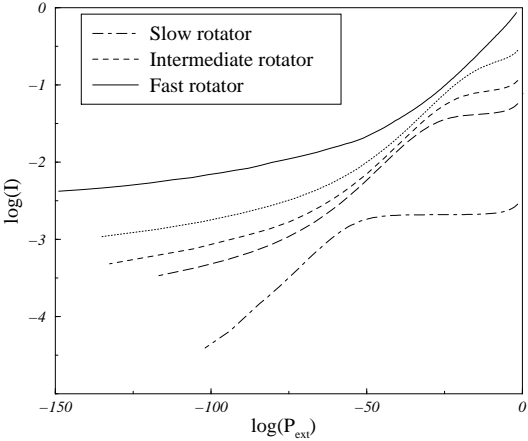}
\caption[ ]{
Variations of the asymptotic electric current with the external 
confining pressure for various types of rigid rotators. 
The solid lines correspond to the fastest rotators while the 
dashed lines stand for slow rotators. $\bar Q=3.1$, 
$\bar \alpha_0=2.5$ and $\bar\Omega$ varies from 0.1 to
15 corresponding to $\omega$ varying from 0.1 to 1.8.
}
\label{fig18}
\end{figure}
%%%%%%%%%%%%%%%%%%%%%%%%%%%%%%%%%%% 
We plot in Fig.~\ref{fig18} the total asymptotic poloidal 
current in the jet as a function of the confining pressure for 
different types of rotators ranging from slow to very fast 
ones. All other parameters are kept constant. The external 
pressure can be reduced until the jet reaches unphysical size. 
The total current diminishes as the external pressure drops for 
all types of rotators but never approaches a constant non-vanishing 
limit  for the smallest values of the pressure. The slowest rotator
in  Fig.~\ref{fig18} shows an effect of pressure threshold. For 
pressures larger than some threshold value the current is almost 
constant but it strongly decreases for smaller ones. For very fast 
rotators  the current varies little with external pressure.

It would have been interesting to reach a conclusion
on whether the asymptotic poloidal current vanishes 
or not in the limit of vanishing confining pressures.
This issue is important because it distinguishes
the different possible asymptotic regimes for
unconfined jets (Heyvaerts and Norman, 1989).
Fig.~\ref{fig18} does not show any leveling-off
of the current $I$ as $log P_{ext}$ approaches $- \infty$. 
Regardless of the smallness of the limiting values of the 
pressure that have been reached, this study does not allow 
to conclude that the current vanishes  when the pressure 
rigorously does. However, from a practical point of view,
it appears that a significant residual current
remains, even for exceedingly  small, but non-vanishing, 
confining pressure.

\subsection{Differential Rotators}

\begin{figure}[htbp]
\psfig{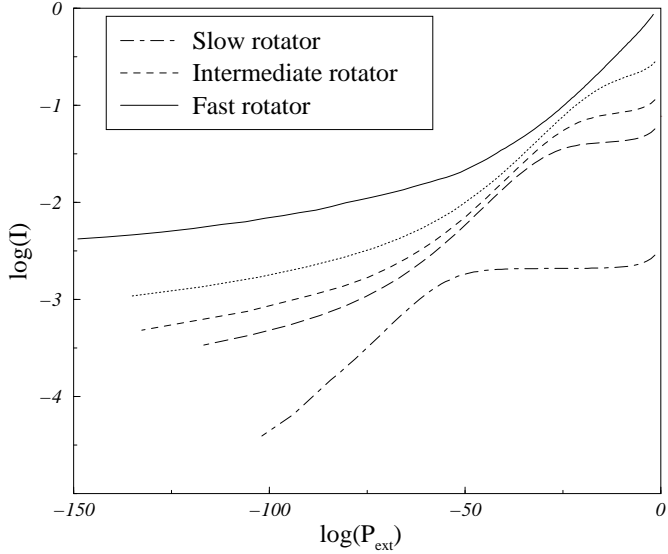}
\caption[ ]{
Plots of the electric current profiles across the flow
with respect to the relative radius $r_{rel}=\frac{r}{r_{jet}}$.
Left panel stands for differential angular velocities and
right panel shows differential specific entropies.
Note the difference of ranges between different panels.
}
\label{fig19}
\end{figure}

We now investigate solutions with large gradients of the angular 
velocity, i.e. $\dot \omega \gg \dot x$. Eq.~(\ref{TRANSgen}) 
can be written as 
\begin{equation}
\left(\gamma-1\right)\beta y^{\gamma-1}\dot y
+2 \omega^2 xy^2 \left( \dot y /y + 
\dot \omega /\omega\right)
 = 0
\end{equation}
In this case the transfield equation can be integrated
neglecting the relative derivative of the radius 
with respect to the relative derivatives of the density
and of the rotation parameter. 
The integral is
\begin{equation}
\frac{\gamma-1}{\gamma}\beta y^{\gamma}
+ \omega^2 x  y^2 = \beta y_0^{\gamma-1}
\label{COM}
\end{equation}
Eq.~(\ref{COM}) can be
transformed using Eq.~(\ref{I1}) into
an equation for  the current, that we will note
$I_{diff}$, the solution of which is
\begin{equation}
I_{diff} = 
\frac{A }{r_A \mu_0}
\frac{\gamma-1}{\gamma}\frac{\beta }{\omega y_b}
\left(y_0^{\gamma-1}-y_b^{\gamma-1}\right)
\end{equation}
This is different from the solution
for the rigid rotator case, that 
can be noted $I_{rig}$. The relation between them is 
\begin{equation}
I_{diff} = 
\left(\frac{\gamma-1}{\gamma}\right)
\frac{2}{y_b} I_{rig}
\end{equation}
This shows that the total current for differential rotators
with large gradients of angular velocity will be larger since 
the density on the boundary is small compared to unity.
This can clearly be seen in the left panel of Fig.~\ref{fig19} 
where the current is plotted with respect to the radius for
different types of differential rotators.  
The largest values of the current correspond to
profiles where the gradients of the angular
velocity are positive and are the largest, namely for
profiles of type $\Omega_2$, where the angular velocity 
takes half of its value at the middle of the outflow.
%%%%%%%%%%%%%%%%%%%%%%%%%%%%%%%%%%%%
%%%%%%%%%%%%%%%%%%%%%%%%%%%%%%%%%%%
Fig.~\ref{fig19} shows that the profile of the angular velocity 
is of prime importance, and that variations of the entropy with flux 
have a lesser influence on the current profile.
Thus {\em jets from differential rotators will carry a larger
current and might be more collimated than rigidly rotating outflows}. 

\section{Comparison between Numerical Results and a Simplified Model}

\begin{figure}[htbp]
\psfig{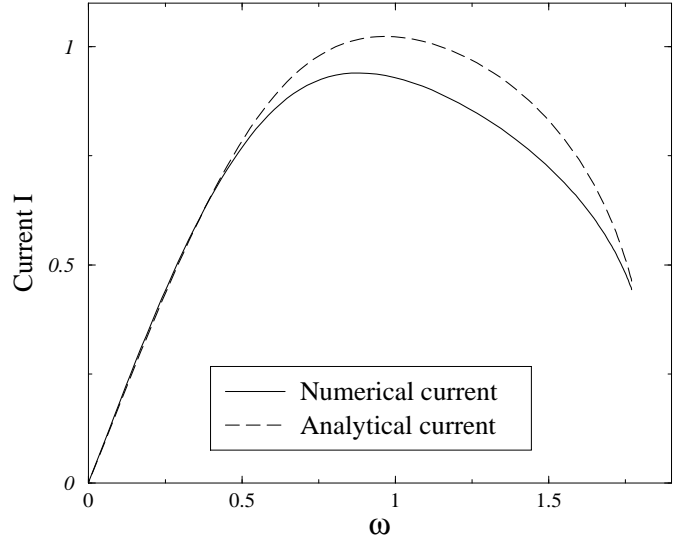}
\caption[ ]{
Comparison of numerical results (solid line)
and analytical solutions (dashed line) of
the asymptotic electric current $I$ 
with respect to the rotation parameter $\omega$.
}
\label{fig20}
\end{figure}

In a review of the theory of magnetically accelerated outflows and 
jets from accretion disks, Spruit (\cite{spruit}) discusses the 
asymptotic wind structure. As in paper I, using his simplifications 
and  method, we find that
\begin{equation}
x y  = 1-\left(\frac{\alpha}{\alpha_*\omega}\right)^{4/3} 
\left(\frac {r_A}{r_*}\right)^2 .
\end{equation}
The net current from this simplified analytical model, 
that we will note $I_{simp}$, is defined by Eq.~(\ref{I1}).
It is given in the present case by
\begin{equation}
I_{simp} = \frac{9 \omega A}{5 \mu_0 r_*}
\left(\frac{ \alpha}{\alpha_*}\right)^{1/3}
\left(1-\left(\frac{5 \alpha}{9\alpha_*\omega^2}\right)^{2/3} 
\right) 
\label{itheo23}
\end{equation}
Thus {\em the current is proportional to the angular velocity
and the mass loss rate with the following dependencies}
\begin{equation}
I_{simp}  \propto \Omega \dot M^{1/3}
\end{equation}
The current can also be given as a function of the rotation 
parameter $\omega$ and of the Alfv\'en radius by
\begin{equation}
I_{simp} \approx 3 \omega^{1/3} A/2 \mu_0 r_A
\end{equation}
We have plotted this analytical solution 
with respect to $\omega$ in Fig.~\ref{fig20}
together with the corresponding numerical solution. 
As $\omega$ increases and passes $0.5$, the two solutions separate 
in the intermediate rotator region to converge again in the limit 
of very fast rotators. As shown previously the current increases
with respect to $\omega$ for slow rotators and diminishes 
in the other limit. So the analytical results we have found for the 
current are in qualitative agreement with  the real value.
%%%%%%%%%%%%%%%%%%%%%%%%%%%%%%%%%%%%
%%%%%%%%%%%%%%%%%%%%%%%%%%%%%%%%%%%

\paragraph{The Poynting Flux}
It is also interesting to calculate the Poynting flux per unit 
escaping mass  which can be written as $S = \Omega r B_\phi/\mu_0 \alpha = 
I \Omega/\alpha$. Using our dimensionless variables it can be expressed as
\begin{equation}
S = \frac{A^2}{\mu_0 \rho_A r_A^4}
\omega^2 x y = \frac{A}{\rho_A r_A^3} I \omega
\label{S1I}
\end{equation}
Using the previous Eq.~(\ref{itheo23}), the Poynting flux for the 
above simplified analytical model is given by
\begin{equation}
S_{simp} = 
\frac{A}{\rho_A r_A^3}
\frac{9 \omega^2 A}{5 \mu_0 r_*}
\left(\frac{ \alpha}{\alpha_*}\right)^{1/3}
\left(1-\left(\frac{5 \alpha}{9\alpha_*\omega^2}\right)^{2/3} 
\right) .
\label{Stheo1}
\end{equation}
Thus {\em the Poynting flux scales with the rotation and the mass 
loss rate as}
\begin{equation}
S_{simp}  \propto \Omega^2 \ \dot M^{-2/3}.
\end{equation}
This analytical solution is plotted with respect to
the relative radius together with the numerical results
with the same input parameters. The agreement is good
all across the outflow, and Eq.~(\ref{Stheo1}) 
well reproduces the behavior of the asymptotic Poynting flux.
Eq.~(\ref{Stheo1}) can be combined with the constraint on the boundary 
mass density $\rho_b$ given in paper I by 
$\left[S^2 H\right]_{slow}\geq \left[S^2 H\right]_{fast}$
where the specific enthalpy $H$ is given at the slow and fast surfaces
Thus the asymptotic density at the outer edge of the outflow
has an upper bound in terms of quantities defined in the inner 
part of the outflow close to its source. 
%%%%%%%%%%%%%%%%%%%%%%%%%%%%%%%%%%%%
\begin{figure}[htbp]
\psfig{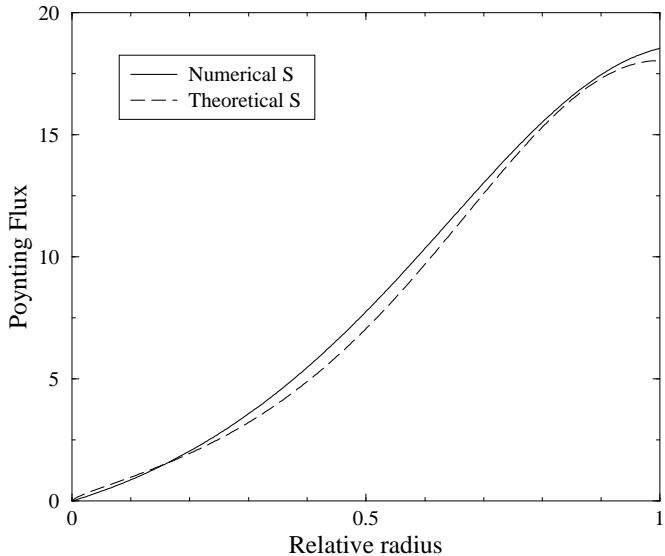}
\caption[ ]{
Comparison of numerical results (solid line) and 
analytical solutions (dashed line) for
the Poynting flux $S$ across the outflow as functions
of the relative radius $r_{rel}=\frac{r}{r_{jet}}$.
}
\label{fig21}
\end{figure}
%%%%%%%%%%%%%%%%%%%%%%%%%%%%%%%%%%
The wind carries both kinetic and magnetic energy, 
the asymptotic ratio of these, at large distance, is a
measure of the importance of the magnetic component.
The kinetic energy flux is $K = \rho v_p^2 /2$.
The Poynting to kinetic energy flux ratio
is given by $q = S/K =3 \omega^{4/3}v_{pA}^2/\rho v_{p}^2$
The effects of the external pressure on the 
Poynting to kinetic energy fluxes ratio $q$  have been
calculated for the different classes of rotators.
It is found that the faster the rotator, the larger
the ratio $q$  which  increases with the confining pressure.
Moreover {\em in the asymptotic regime, most of the magnetic 
energy has been transformed into kinetic energy in the central 
core close to the axis, contrary to the envelope where $q$ can 
be very large}.

\section{Conclusions }

We have considered the asymptotic behavior of outflows of magnetized 
rotators confined by an external uniform pressure in the framework of 
our simplified model. The  asymptotic forms of the transfield and the 
Bernoulli equations were used to determine the jet structure  taking 
into account the pressure balance across the interface between the flow 
and the external confining medium. No self-similar assumption has been 
made. The given confining pressure has been regarded as a boundary 
condition and the constants of the motion obtained in the inner part 
of the flow close to the emitting source have been used. The full 
range of possible variations of the parameters has been explored. 

Slow rotators are dominated by thermal effects from the axis to the 
outer edge. The specific entropy however has little influence on 
their asymptotic magnetic field. In the case of fast rotators rotational 
and mass loss rate effects have the most important influences on 
solutions. Magnetic pressure dominates at the outer edge of the flow. 
Such rotators asymptotically have a concentrated azimuthal magnetic 
field $B_\phi$, small mass loss rates and large density gradients. 
A large part of the magnetic energy is not transferred to the kinetic 
energy and the outflow is asymptotically strongly magnetized  and 
carries a significant Poynting flux. 

Whatever the type of rotator, isothermal jets are narrower than 
adiabatic ones under the same conditions. The densest jets are 
slower on the boundary than lighter ones.  In the case of slow 
rotators,  an analytical solution for the current in terms of the 
axial density, thermal and rotation parameters has been obtained. 
This relation combined with the analytical solution gives the 
asymptotic current as a function of the confining pressure. 
An analytical solution for the poloidal current has also been 
obtained in the case of the fast rotator in terms of the rotation
 parameter and the Alfv\'en radius. The current in slow rotators 
is  diffuse in the outflow while fast rotators  carry a concentrated 
electric current around the axis. The solutions obtained with our 
model can carry return currents out of the polar axis, contrary to 
self-similar models. The comparison between numerical results and 
approximate analytic solutions show  the latter to be good qualitative 
estimators of the real value.

Non constant profiles of rotation and, to a lesser extent of entropy,   
cause the solutions to change drastically. For example, the largest 
asymptotic poloidal velocity can  be located either on the axis or 
at the outer edge according to the profile of $\Omega(a)$. It has been 
possible to find solutions resembling observed flows such as central 
jets with important momentum surrounded by a larger outflow. Large 
currents can be generated by  differential rotators if the gradient of 
the angular velocity is large and the angular velocity does not vanish 
on the outer edge of the flow.

Thus  our model makes it  possible to relate the properties of the 
asymptotic part of an outflow to those of the source. These asymptotic 
equilibria can be used as input solutions for numerical simulations 
in order to investigate the propagation of jets and the instabilities 
that can develop in magnetized outflows.

\begin{acknowledgements}
We would like to thank Kanaris Tsinganos for his remarks as referee
that helped to clarify some of the derivations and discussions
of equations.
\end{acknowledgements}

\end{document}